\documentclass{cernrep}
\begin{document}
\title{Axion-like Particles from Primakov production in beam-dumps}
\author{Babette D\"obrich}
\institute{CERN, 1211 Geneva 23}

\begin{abstract}
We discuss searches for Axion-like Particles which are coupled predominantly to
photons from proton- or electron beam-dumps.
In particular, we scrutinize the present state of exclusions from SLAC 141 
in the mass range of $\sim$ 10-30~MeV
\end{abstract}


\maketitle

\section{ALPs in the MeV-GeV mass range}

Whilst we have good reason to believe in the existence of particles
beyond the Standard Model, there are many well-motivated
but also very different places to look for them:
In the sub-eV range, e.g. the QCD axion, which is produced non-thermally in the early
universe, is an excellent candidate for Dark Matter.
At masses around $\sim$ GeV, WIMPs can reveal themselves by scattering in ultra-low background detectors
or be seen indirectly at the LHC, see e.g. \cite{Baudis:2015mpa}. 
The (so-far) non observation of new particles suggests to probe those
energies also under a different view-point:
The particles that could be Dark Matter may not couple directly to the Standard
Model but through a weakly coupled `portal', for which different possibilities and motivations exist, 
see \cite{Alekhin:2015byh} for a review.

In the following, we review some physics results and prospects for  a pseudo-scalar Axion-like Particles (ALP)
somewhat below the GeV scale. Such an ALP can act as portal \cite{Berlin:2015wwa} and
can in specific cases be even useful to solve the strong CP problem \cite{Berezhiani:2000gh}.

If such an ALP $a$ is predominantly coupled to photons we can write the interaction as:
\begin{equation}
\mathcal{L}= \frac{1}{2} \partial^\mu  a \, \partial_\mu a - \frac{1}{2}m^{2}_{a} \ a^2-\frac{1}{4} \  g_{a \gamma}  a \, F^{\mu\nu}\tilde{F}_{\mu\nu} \; ,
\label{eq:lagr}
\end{equation}  	
where $g_{a \gamma}$ denotes the photon-ALP coupling. A powerful method to search
for them is in proton- or electron beam-dumps. 
In such experiments, ALPs can be created  from Primakov production.
The advantage in ALP production from coherent proton- or electron-scattering
is that the production cross-section can be reliably computed. Particularly the
transverse momenta of the ALP which are crucial
to be known for an accurate estimate of the ALP acceptance in a beam-dump setup. 
Additionally, ALP Primakov production in the dump can be boosted
with heavier target nuclei. In \cite{Dobrich:2015jyk}
details of computing the differential cross-sections (with respect to energy
and transverse angle)
of such ALPs have been worked out. Sensitivity estimates
for past and future proton-beam dumps have been provided and
compared to the literature on electron-beam dump results.

In the following, let us use the opportunity to re-visit the electron
beam-dump limit put on ALPs from SLAC 141. This is particularly interesting
as upcoming and planned experiments can partially overlap with the parameter space 
excluded by SLAC 141.

\section{Limits on ALPs from beam-dumps}

\begin{figure}[ht]
\begin{center}
\includegraphics[width=8cm]{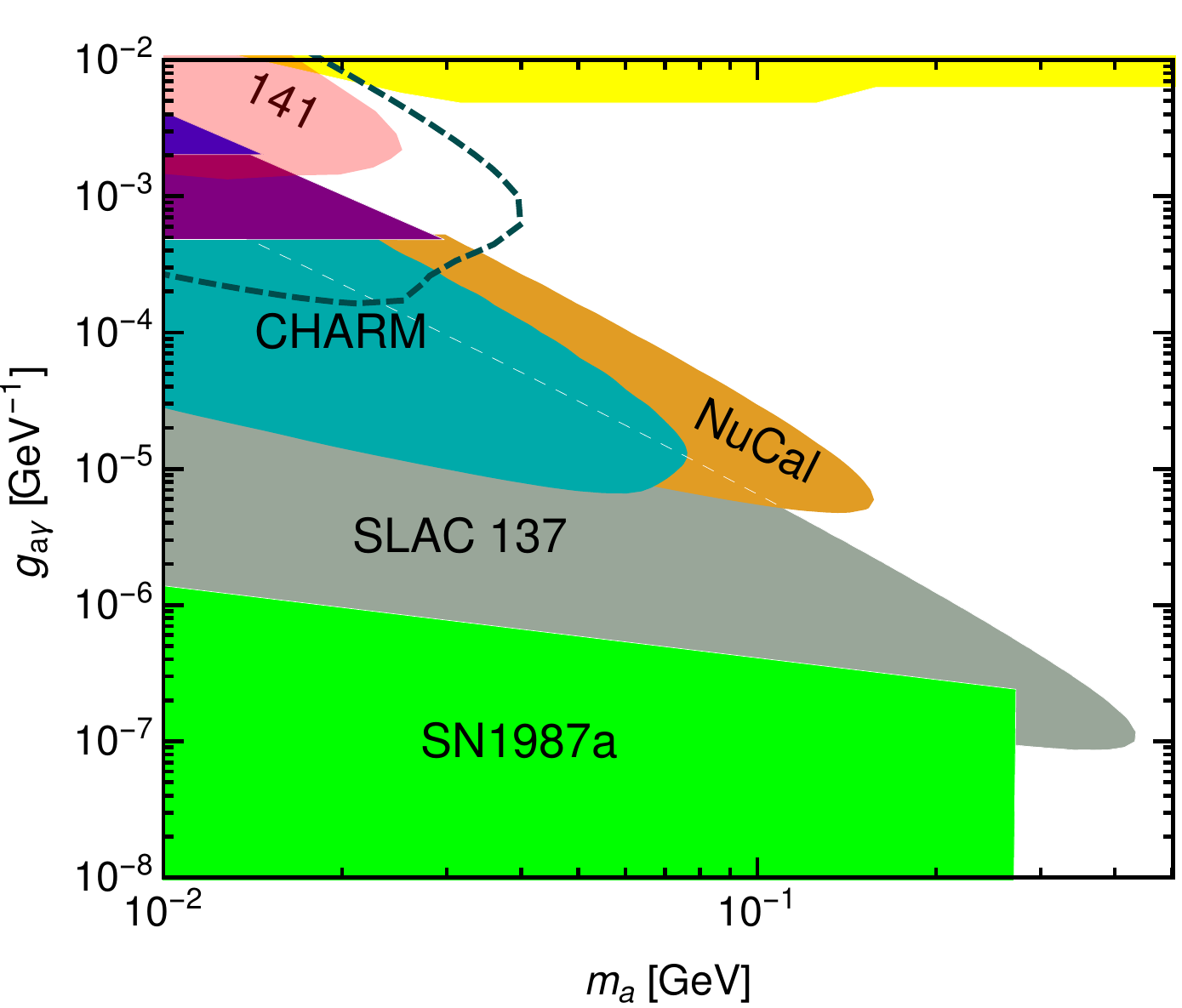}
\caption{Compilation of limits on ALPs coupled to photons. All limits are shown at 90\% C.L. .
The compilation is the same as in  \cite{Dobrich:2015jyk} except for the limits set by SLAC 141:
The dark-green dashed line
shows its exclusion inferred from a negative result of a $e^+, e^-$ search.
The rose region shows exclusion based on an analysis by the collaboration of the same data set.}
\label{fig:Limits}
\end{center}
\end{figure}

Figure  \ref{fig:Limits} (adapted from \cite{Dobrich:2015jyk} w.r.t. the SLAC 141 region, as
explained in the text below) summarizes presently published exclusions on ALPs in the MeV-GeV mass range.
The shape of the regions excluded by individual experiments can be understood
as follows: The ALP of Eq. \ref{eq:lagr} has a lifetime 
\begin{equation}
 \tau  =  64 \pi/ (g_{a \gamma}^2 m_a^3) \ .
 \label{eq:lifetime}
 \end{equation}
 The requirement that ALPs decay after the beginning of the fiducial region 
 but before its end yield the characteristic nose-shape of the beam-dump exclusion
 regions. Let us describe these regions in the following.

In Figure  \ref{fig:Limits},
`CHARM' and `NuCal' are exclusions
set from $\sim 2.4 \times 10^{18}$ and $\sim 1.7 \times 10^{18}$ protons on target, respectively. Note, that albeit 
NuCal had a lower beam energy of 70 GeV than CHARM (400 GeV), the NuCal detector was much closer (64m) to the production point
than CHARM (480m)
which allows the search for ALPs of much shorter lifetimes. On the other 
hand, `SLAC137' is based on $2 \times 10^{20}$ electrons
on an aluminum target 200~m upstream of the detector.
Details of the limit compilation, references to the experiment papers and
a description of the other exclusion regions can be found in \cite{Dobrich:2015jyk}.

In previous versions of Figure \ref{fig:Limits}, limits from SLAC 141 on ALPs have been shown based on reported
negative search results \cite{Riordan:1987aw} of ALPs decaying to $e^+, e^-$. The experiment used an incident beam of 
9 GeV and the
spectrometer, located 35m downstream the dump, looked for positrons of energies of minimum 4.5~GeV.  
This data taking included also
the regular insertion of a photon converter such that it had sensitivity to di-photon final states.
In the following, we point to a less-known analysis of this data for ALP $\rightarrow \gamma, \gamma$
and therewith update the status of exclusion shown previously.

The dark green, dashed line in Figure \ref{fig:Limits} is the SLAC 141 limit on $g_{a \gamma}$ shown
first in a review article of a 2011 community workshop \cite{Hewett:2012ns},
referencing the SLAC 141 publication \cite{Riordan:1987aw}. The publication \cite{Riordan:1987aw} reports on a search for light,
pseudo-scalar bosons, predominantly coupled to $e^+,e^-$. Based on $\sim 2 \times 10^{15}$
electrons on a tungsten target they excluded such ALPs up to
masses of $\simeq 20$MeV. Part of this data was taken with a photon converter in place
as pointed out in the letter.

The rose-shaded labeled `141' in Fig.~\ref{fig:Limits} presents the exclusion limit from Figure 3 of \cite{krasny}.
Ref \cite{krasny} is based on the same SLAC 141 data
and presents a  limit for an ALP coupled to two photons.
To translate the results presented in \cite{krasny} into Figure \ref{fig:Limits},
we have employed a conversion of the ALP lifetime provided in Eq.~\ref{eq:lifetime}.

As is apparent, the limit on $g_{a \gamma}$ based on the results of \cite{Riordan:1987aw} seem to be much stronger than the limits
reported from the di-photon search, albeit both limits are referring to the same data.
The reason for this discrepancy can possibly be understood as follows: As mentioned above,
the search for ALPs coupled to two photons was made possible by the insertion of a photon converter.
To infer an exclusion limit on ALPs coupled predominantly to photons from the information in \cite{Riordan:1987aw},
however requires the knowledge of the spectrum of the charged particles from the photon converter:
Since the cut of 4.5~GeV on the secondary $e^+$ is so close to the beam energy of the dumped beam, the yield is very
sensitive to the required minimum ALP energy.
It is therefore a non-trival task to infer limits $g_{a \gamma}$ solely based on information stated in the $e^+,e^-$ search.

The publication of \cite{krasny} is - to our knowledge - not easily found on-line, but it is the relevant reference for
limits on $g_{a \gamma}$ from SLAC 141.
We conclude that for future experiments it can be worth searching the `white region' below
the dash-dotted, green line of Figure \ref{fig:Limits}.

\section{Future searches for ALP at MeV-GeV}

Current and future searches have the prospect of searching the `blank space' of figure \ref{fig:Limits}:
The fixed target experiment NA62 \cite{NA62:2017rwk} is currently taking data 
to measure $K^+\rightarrow \pi^+\nu \bar{\nu}$, 
but is in principle able to run in beam-dump mode:
Removing the target and `closing' the (copper) collimator can open up significant discovery potential
for ALPs \cite{Dobrich:2015jyk}.
For a discovery of ALPs, at minimum an analysis for which both photons from the decay
are detected, is needed.
If tracking of the photons is not possible, searching for any
`Dark Particle' decaying into two photons can be challenging 
since the decay point cannot be determined when the mass of the decaying particle is unknown.
However, it {\it is} possible to reliably
compute the angular distributions of ALPs created from coherent Primakov production.
It is useful to exploit that information to single out a potential ALP-signal.

In the near future, depending on the model, e.g. Belle2 \cite{Izaguirre:2016dfi},
the proposed SHiP experiment \cite{Anelli:2015pba} and the LHC
\cite{Knapen:2016moh,Bauer:2017ris} have excellent potential of discovering 
ALPs coupled to photons in that mass range.

In summary,
checking for the existence of weakly coupled particles at the MeV-GeV scale 
deserves dedicated effort. 
Compiling accurate existing constraints is a
crucial input for new searches.

\section*{Acknowledgements}

I wish to thank M.~W.~Krasny for bringing the SLAC 141 results presented at EPS 1987 to our attention.
I am grateful to Felix Kahlhoefer for helpful comments on this proceedings article.
Also, I would like to thank the organizers of PHOTON 2017 for bringing this great workshop to CERN.

\end{document}